# Oxygen in diamond: thermal stability of ST1 spin centres and creation of oxygen-pair complexes

Paul Neugebauer[1], Xinxi Huang[1,2], Chloe Newsom[3], Christophe Arnold[4], Hjørdis Martelock[1], Séverine Diziain[1], Edoardo Monnetti[1,2], Jocelyn Achard[5], Tobias Lühmann[1], Paolo Olivero[2], Jan Meijer[1], Julien Barjon[4], Alexandre Tallaire[3] and Sébastien Pezzagna[1,*]

[1]Applied Quantum Systems, Felix-Bloch Institute for Solid-State Physics, Universität Leipzig, Linnéstraße 5, 04103 Leipzig, Germany

[2]Physics Department, University of Torino, Torino 10125, Italy; Instituto Nazionale di Fisica Nucleare (INFN), Sezione di Torino, Torino 10125, Italy; Instituto Nazionale di Ricerca Metrologica (INRiM), Torino 10135, Italy

[3]IRCP, CNRS, PSL Research University, 11 rue Pierre et Marie Curie, 75005 Paris, France

[4]LSPM-CNRS, UPR 3407, Université Sorbonne Paris Nord, 99 Avenue J. B. Clément, Villetaneuse F-93439, France

[5]GEMaC-CNRS/USVQ, Université Paris-Saclay, Versailles, France

*Sebastien.pezzagna@uni-leipzig.de

## Abstract

Little is known about oxygen-related defects in diamond. Recently, the promising room-temperature spin centre named ST1 was identified as an oxygen centre, but of still unknown atomic structure and thermal stability. In this work, we report on the optically active oxygen-related centres and the conditions for their formation, using ion implantation of oxygen in various conditions of depth and fluence. More specifically, we establish the temperature formation/stability range of the ST1 centre, which has a maximum at about 1100°C and is narrower than for NV centres. In these conditions, optically detected magnetic resonance (ODMR) on small ST1 ensembles was measured with a spin readout contrast of > 20% at 300K. In cathodoluminescence, the 535 nm ST1 peak is not observed. Besides, a broad peak centred at 460 nm is measured for implantation of $O_2$ molecular ions. For an annealing temperature of 1500°C, a different centre is formed (with ZPL at 584.5 nm) with an intensity increasing with a power law $1.5 < p < 1.9$ dependence from the implantation fluence. This suggests that this centre contains two oxygen atoms. Besides, a new spectral feature associated to an intrinsic defect was also observed, with four prominent lines (especially at 594nm). Finally, the thermal formation and stability of oxygen centres in diamond presented here are important for the identification of the atomic structure of defects such as the ST1 and possible $O_2V_x$ complex by means of *ab initio* calculations. Indeed, the formation energies and charge states of defect centres are easier to compute than the full energy level scheme, which to date still remains unsuccessful regarding the ST1 centre.

**Introduction**

Defect centres in diamond such as the Nitrogen-Vacancy (NV) centre paved the way to quantum sensing applications [1,2]. Many of these defects have been widely studied for years but little is known about oxygen-related defect centres, and this mainly for two reasons. The natural abundance of the three stable oxygen isotopes are 99.76% for $^{16}$O (I=0), 0.20% for $^{18}$O (I=0) and only 0.04% for $^{17}$O (I=5/2), which makes any detection of the nuclear spin highly improbable. Secondly, oxygen is an etching species in the CVD growth of diamond [3] (which is the method of choice for high purity samples), leading to volatile CO species, and therefore oxygen does not easily incorporate into the crystal lattice. It was calculated by ab initio methods [4] that the preferential form for isolated oxygen in diamond is substitutional rather than interstitial, with a significant difference in formation energies. The EPR defect named KUL12, which was found after oxygen implantation and thermal annealing [5], is formed quite efficiently upon ion implantation, i.e. about 8% of the implanted oxygen atoms are in this form. Furthermore, annealing up to 1400 °C does not strongly modify its concentration [5]. This defect has been attributed to $O_s^+$ by Thiering & Gali [4]. Substitutional oxygen can also be present in the neutral charge state, but it is not measurable by EPR. The value of 8% is therefore an underestimation of the total $[O_S]$ concentration. Besides, the neutral oxygen-vacancy $OV^0$ was for some time expected to be a promising qubit candidate [6] because of its similarity to the negatively-charged NV$^-$ complex. However, it is characterised by a strong non-radiative decay channel and was not observed in PL, although a ZPL possibly at 543 nm is discussed in [7]. The presence of the vacancy in the $OV^0$ centre was proved by increasing the concentration of oxygen-vacancy complexes with the introduction of additional vacancies followed by annealing [8], in analogy with what has been observed with NV centres [9]. Considering ion implantation, with large (i.e. > $10^{15}$ cm$^{-2}$) oxygen fluences followed by high (i.e. 1600 °C) temperature annealing, a broad PL emission related to oxygen was also reported, with characteristic lines at 584.5 nm and 598.4 nm [10]. This was found also by Lühmann et al. [11] and attributed to an oxygen complex.

On the other hand, the so-called ST1 centre [12-15], which is a promising spin centre for quantum sensing and nuclear spin memories at room temperature, was very recently proved to contain oxygen. Remarkably, its atomic structure is still unknown and it presents the particularity to possess (at least) four stable configurations, usually referred as polymorphs, which are likely due to a pseudo Jan-Teller relaxation of the O atom within the structure [16]. The electronic structure of these ST1 centres consists of singlet ground and excited states with a metastable spin triplet accessible through intersystem crossing. The ZPL of the main polymorph is at 535.1 nm and the zero field splitting $D$ and crystal field parameter $E$ are 1133.5 MHz and 143.5 MHz, respectively, at room temperature. Interestingly, ST1 can be produced very reproducibly despite its extremely low formation yield of a few $10^{-5}$ [16]. Attempts to identify the atomic structure of the ST1 centre by *ab initio* calculations were, until now, not successful: the computed electronic structure and energies of several oxygen-related defects [4,17] such as OV, OCV, OCCV or OCCCV do not match the experimentally observed properties of the ST1 centre. The main reasons for this can be identified in the strongly correlated orbitals of oxygen, the relatively large extension of the ST1 electronic wavefunction [12,16] regarding the size of the supercells used, and the highly time-consuming process to calculate the electronic structure. On the other hand, formation energy is a much easier property to calculate. It is therefore one of the purposes of this work to determine the temperature formation range of ST1 centres, in a set of different conditions such as depth (i.e. Fermi energy, because of band bending close to the diamond surface) and oxygen concentration. By combining photoluminescence (PL) and cathodoluminescence (CL) experiments at room-temperature, we present a spectroscopic study of optically active defect centres related to oxygen in diamond and discuss how they form or dissociate. Furthermore, the PL *vs* CL results give insights into the different charge states and excitation pathways of the centres.

**Ion implantation of oxygen centres in diamond**

Three diamond samples of type IIa with (001) surface orientation were used: an as-grown (non-polished) sample made at LSPM laboratory (sample #1) and two quantum-grade diamond sample from Element6 with polished surface (samples #2 and #3). The native nitrogen concentration of sample #1 is estimated to be about 0.05 ppb (assuming a [NV]/[N] ratio of 1/500) with respect to the measured native NV concentration of about $1.7×10^{10}$ cm$^{-3}$. For samples #2 and #3, it is lower than 5 ppb according to the producer's specifications. Before implantation, the surface of sample #1 is H-terminated while it is O-terminated for samples #2 and #3, which can have an effect on defect creation [18,19]. The three samples were implanted with oxygen ions, at a range of energies (from 5 keV to 80 keV) and fluences (from $1×10^{12}$ cm$^{-2}$ to $1×10^{15}$ cm$^{-2}$). Sample #1 was implanted at IRCP with a focused ion beam (FIB) and samples #2 and #3 were implanted at University Leipzig with a 100 kV accelerator (see Experimental details).

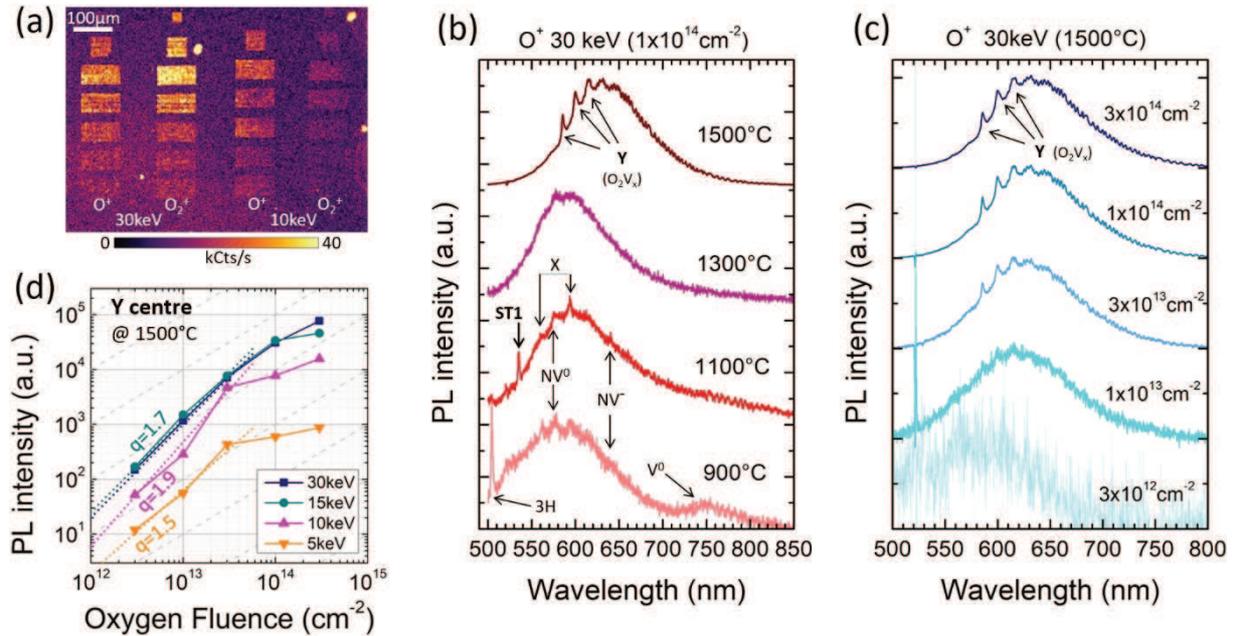

**Figure 1. Annealing temperature dependence of PL spectra in sample #1.** (a) PL scan of sample #1 (as grown) after ion implantation and the first annealing at 900°C for 4 hours. The laser excitation is 488 nm and the detection window 580 – 635 nm. (b) PL spectra recorded in the area 30 keV, O$^+$, $1×10^{14}$ cm$^{-2}$, after each annealing step. The spectra were recorded by scanning over the implanted area, for 60 seconds. The spectra are background corrected, normalised to their maximum and separated for clarity. (c) PL spectra of the 30 keV, O$^+$ implantations of different implanted fluences, for the 1500°C annealing. The spectra were recorded by scanning over the implanted area, for 60 seconds. The spectra are background corrected, normalised to their maximum and separated for clarity. (d) PL intensity of the spectra at 1500°C (same window as in (a)), attributed to an oxygen-pair complex, as a function of the oxygen fluence, for the different ion species and energy per oxygen atom. The grey dashed lines would represent the linear dependence between PL intensity $I_{PL}$ and ion fluence $F$. The dotted lines are fits for the regime below saturation (< $3×10^{13}$ cm$^{-2}$) revealing a superlinear dependence $I_{PL} \propto F^q$, with 1.5 < q < 1.9 indicated on the graph.

**Temperature evolution (sample #1)**
Sample #1 was dedicated to the annealing temperature investigations and formation of the different centres. After ion implantation, the sample was annealed in vacuum over four sequential 4-hours runs at increasing temperature (namely: 900 °C, 1100 °C, 1300 °C and 1500 °C) and optically characterised between each step. A PL scan of the sample after the 900 °C annealing is shown in Figure 1a. The different implantation fields can be clearly identified. In Figure 1b, PL spectra measured in the area implanted with O$^+$ ions at 30 keV and a fluence of $1×10^{14}$ cm$^{-2}$ are shown as a function of the annealing temperature. At 900 °C, the main features are the 3H interstitial defects (504 nm), V$^0$ neutral individual vacancies (741 nm) but also NV centres (575 nm for NV$^0$ and 638 nm for NV$^-$). The presence of NVs is due to the vacancies created upon ion implantation, binding with native nitrogen impurities and forming NV centres during the annealing. At this temperature, no significant presence of ST1 centre can be noted. At 1100 °C, the 3H and V$^0$ peaks are strongly reduced. The NVs are still present and a clear ZPL at 535nm of the ST1 centre is now visible. It is worth noting the presence of features at 562 nm and 595 nm, noted as "X", which will be discussed later with CL experiments. At 1300 °C, the ST1

ZPL has almost vanished whereas the broad band with a maximum at about 590 nm is still present. It is not clear whether all ST1 centres have disappeared or if a redistribution of the polymorphs on the other emission wavelengths [16] has also contributed to the loss of the 535 nm ZPL intensity. Subsequently, a new centre forms upon 1500 °C annealing, as already observed [10,11] (labelled here as "Y") with a broad peak and prominent narrow lines at 584.5 nm and 598.5 nm. This same characteristic spectrum is observed for all implantation energies (corresponding to different formation depths), but with different intensities. In Figure 1c, the PL spectra acquired from the $O^+$ implantation at 30 keV and annealing at 1500 °C are shown for the different implanted oxygen fluences. The spectra are similar, apart from the lowest fluence of $3\times10^{12}$ cm$^{-2}$. A systematic analysis of the intensity of each spectrum as a function of fluence and energy is shown in Figure 1d. It is worth remarking that in the lower fluence regime (i.e. $\leq 3\times10^{13}$ cm$^{-2}$), the PL intensity does not increase linearly with the oxygen fluence $F$ but rather with a $F^q$ power law, with $1.5 < q < 1.9$. As the fluence increases, the average O – O distance decreases; this effect enables, for the 1500 °C annealing temperature, the formation of the Y centre through oxygen diffusion. Note that the Y centre is not observed upon 1300 °C annealing, even for the highest oxygen fluences. This super-linear (almost quadratic) dependence suggests that the Y colour centre is made of two oxygen atoms, contrarily to the ST1 which shows a linear dependence [16] and is expected to involve only one oxygen atom. We tentatively attribute the Y centre to $O_2V_x$. At fluences larger than $3\times10^{13}$ cm$^{-2}$, a saturation regime of the PL intensity takes place which is observed for most of other defect centres in diamond and likely due to a too high defect concentration. We remark that the onset fluence for PL saturation increases with implantation depth; this effect is expected, because the local implanted ion and vacancy concentration (for a given fluence) decreases with ion energy [11] (see also Figure 3a). Besides, the data in Figure 1d show a clear dependence of the PL intensity of the Y centre on the implantation energy. This is likely due to the band bending at the surface leading either to different creation efficiencies of the Y centre depending on its depth or on different charge state population. An analogy can be noted here with regards to the NV → NVN (H3) centres conversion when the nitrogen-nitrogen average distance is small enough and the annealing temperature high enough to enable nitrogen diffusion [11,20]. By scanning around the implanted area, it was however impossible to isolate such centres at the single level. This indicates that they are weakly radiative, and that they are formed in a significant amout with respect to ST1 centres because the PL intensity of the ensemble is large.

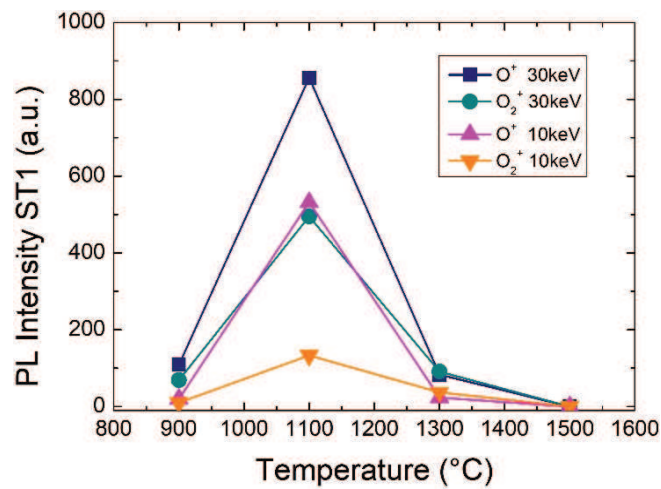

**Figure 2. Thermal stability of ST1 centres.** PL intensity of the ZPL of ST1 centres as a function of the annealing temperature, for different ion species and energies, at a fluence of $1\times10^{14}$ cm$^{-2}$. Each annealing step was done for 4 hours in vacuum.

From the set of measurements carried on sample #1, the temperature range of the ST1 stability can be deduced. To this purpose, the ZPL intensity at 535 nm is plotted in Figure 2 as a function of the annealing temperature and for different ion depths, for a $1\times10^{14}$ cm$^{-2}$ fluence. The data present a maximum at 1100 °C with almost no ST1 at 900 °C or 1300 °C. This behaviour is drastically different from the NV centre or other impurity-vacancy centres in diamond. For example, NV centres already form efficiently at 800 °C and even at slightly lower temperatures [18], without significant improvement at higher temperature [21]. We expect that the main O-related defects in diamond are $O_s$, OV and OVH and that they already form below 900 °C. None of them seems to be visible in the spectra of Figure 1. The temperature range for the existence of ST1 in Figure 2 provides important information on its atomic structure, which noetheless requires a theoretical support to be fully understood. Likely, the ST1 has a more complex structure, obviously less probable to form than O, OV or OVH, because its creation yield was reported to be about $10^{-5}$ [16]. This specific aspect will be investigated in more details in the following with sample #2.

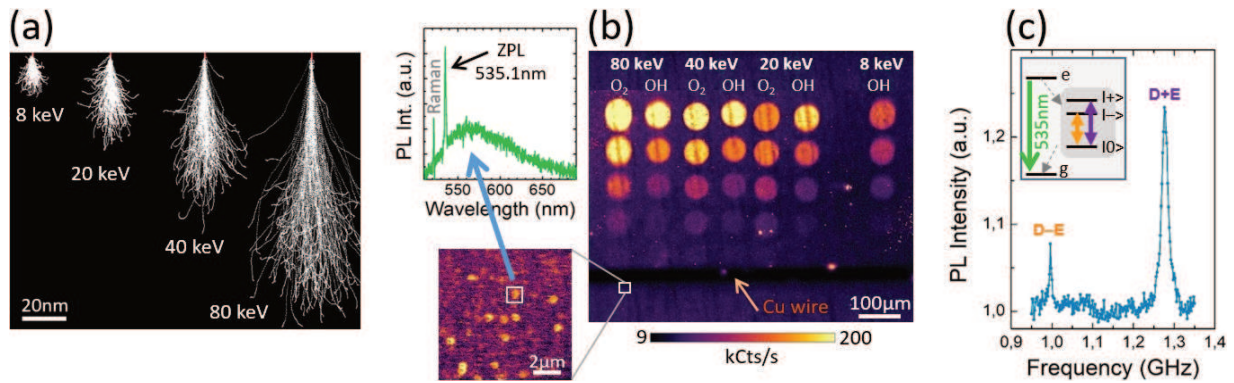

**Figure 3. Ion implantation of oxygen centres in diamond sample #2.** (a) SRIM simulations of the trajectories of 100 oxygen ions implanted at the energies used for sample #2. (b) Confocal fluorescence scan of the implanted area of sample #2 after annealing at 1100°C for 4 hours, revealing ensembles of ST1 centres. A 488nm argon laser is used, with a bandpass filter 580 – 635 nm for imaging. The dark horizontal line is due to a copper wire antenna which was soldered to address the centres with microwaves and identify ST1 through optically detected magnetic resonances (ODMR). The inset (bottom) is a closer scan (10×10 µm²) taken in the lowest implanted area ($10^{12}$ cm$^{-2}$), revealing single ST1 centres. The inset (top) shows the PL spectrum of a single ST1 centre of the main polymorph with ZPL at 535.1 nm. The laser excitation is 488 nm, with 600 µW power. (c) Optically detected magnetic resonance (ODMR) spectrum taken at 300 K in a small ST1 ensemble showing the D-E and D+E transitions within the metastable spin triplet.

**Competing centres at optimal ST1 annealing temperature (sample #2)**

With sample #2, ST1 centres were created by ion implantation and thermal annealing at the optimised temperature of 1100 °C. Figure 3a reports the result of a TRIM simulation [22] showing the trajectories of oxygen ions in diamond for the different energies used for sample #2. Neglecting ion channelling, these respectively correspond to average implantation depths of 11 nm, 25 nm, 48 nm and 91 nm. After annealing, the sample was placed for 30 min in an oxygen plasma cleaner apparatus. As very recently discovered, the ST1 centre exists in different polymorphs, the most abundant one being the one characterised by a ZPL emission at 535.1 nm [16]. To excite them, a 488 nm laser is suitable. However, it is interesting to excite the ensembles with a 532 nm laser, to possibly identify different charge states and compare with previous results from the literature (532 nm excitation was generally used). Figure 3b shows the PL scan of the implanted region using confocal fluorescence microscopy. A 488 nm laser excitation and a 582 – 636 nm detection window were adopted, covering about 40% of the ST1 spectral range of emission. Small ensembles of ST1 centres are produced for the highest oxygen fluences, whereas for the lowest ones a scarce distribution of individual ST1 centres can be seen (inset of Figure 3b).

ODMR measurements were carried both for a small ST1 ensembles and at the single centre level. In Figure 3c, the continuous-wave ODMR spectra at zero external magnetic field corresponding to the OH spot at 80 keV and 1×10$^{14}$ cm$^{-2}$ is shown. The scan range was set in order to detect both the $|0\rangle \leftrightarrow |-\rangle$ and $|0\rangle \leftrightarrow |+\rangle$ electronic spin triplet transitions, also named (D − E) and (D + E) transitions,

respectively. The contrast values of the D – E and D + E peaks are here about 8% and 24%, respectively. These can be further improved using a higher laser excitation power, resulting in more time spent in the spin triplet than below saturation. For ensemble-based magnetometry purposes, the ST1 centres can be found in six possible directions along the [110] axis, whereas the NV has four possible orientations along [111] and the TR12 defect [23,24] has twelve possible orientations. Interestingly, the ODMR contrast of the ST1 centre in the presence of a perpendicular magnetic field does not vanish [16] as is the case for the NV centre, which can be of interest for some applications.

The dependence of the PL spectra on the ion fluence and ion energy was investigated. Figures 4a and 4b show the spectra of the same ensembles ($O_2^-$, 80 keV) excited at 532 nm and 488 nm, respectively. Interestingly, the obtained spectra are very different depending on the excitation wavelength. With 488 nm excitation, the ST1 ensembles present the same feature, i.e. a characteristic ST1 spectrum with variable intensity, depending on the ion fluence or implantation energy and with a more or less pronounced ZPL at 535.1 nm. With 532 nm excitation, the spectra show additional features. At first, a broad peak centred at about 740 nm appears, of comparable intensity with the ST1 fluorescence. This peak is attributed to neutral individual vacancies $V^0$ (also named GR1 with ZPL at 741 nm). The absence of ZPL is attributed to local stress.

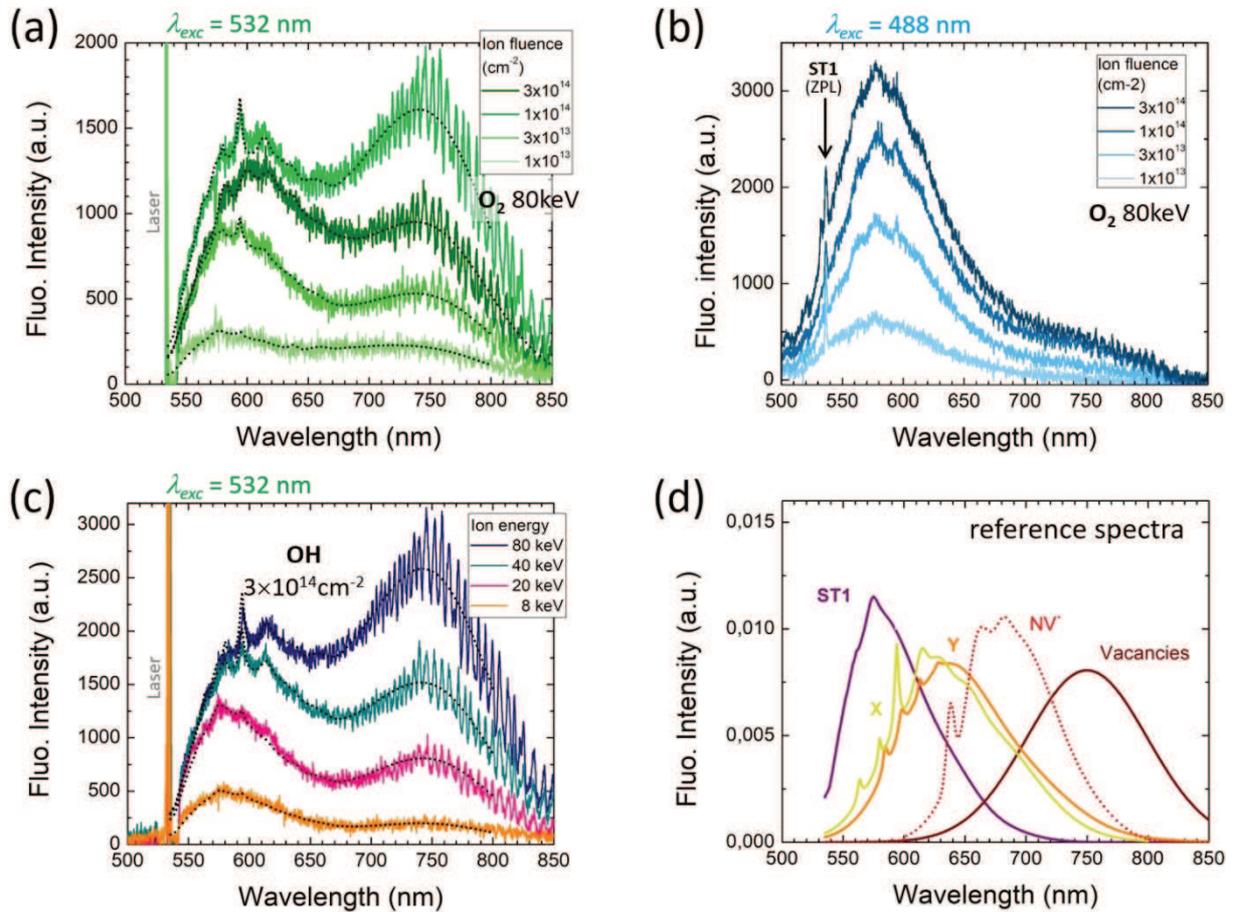

**Figure 4. Spectral properties of ST1 ensembles.** (a) Fluorescence spectra of ST1 ensembles created by $O_2^-$ ions at 80 keV and different fluences, with 532 nm laser excitation. The dotted lines are fitting curves obtained with Equation 1. (b) Spectra of the same implantation areas but recorded with 488 nm excitation. (c) Fluorescence spectra of the $3\times10^{14}$ cm$^{-2}$ OH$^-$ areas implanted at different energies, under 532 nm excitation. The dotted lines in Figures 4a and 4c are fitting curves obtained with Equation 1. (d) Reference spectra used to fit the experimental data as a linear combination of these. The spectra are normalised so that their integrated intensity is equal to 1. The ST1 reference spectra is the average spectrum of three single ST1 centres excited with 532 nm, therefore the 535 nm ZPL is not visible, being cut by the Notch filter (10 nm width) used to remove the laser line.

Furthermore, a slight apparent shift of the ST1 phonon sideband towards longer wavelengths seems to occur, when the ion fluence is increased (Figure 4a) or when the ion energy becomes larger (Figure 4c). This apparent shift can however be explained by the appearance of a broad feature with a maximum at about 630 nm that could be due either to another charge state of ST1 or to another defect formed at high fluence. Note that at 1100 °C the Y centre does not form (see Figure 1b) and cannot be the reason for it. What is occuring here is the formation of the X centre, as will be confirmed later by CL measurements. The prominent peak at 594.5 nm of the X centre appearing in Figure 1b is also visible in the spectra of Figure 4 for the highest implantation depths and fluences.

To account for these different defects, the experimental spectra taken at 532 nm are fitted with the following function,

$$f_{total}(\lambda) = a\, f_{ST1}(\lambda) + b\, f_X(\lambda) + c\, f_{NV}(\lambda) + d\, f_{vac}(\lambda) \tag{1}$$

where $f_{ST1}$, $f_X$, $f_{NV}$ and $f_{vac}$ are area-normalised reference spectra of ST1 (taken with 532 nm excitation, therefore without the 535 nm ZPL), unknown defect named X (see last section), NV centre and vacancies, respectively (Figure 4d). Adding the X centre spectrum enables a very good fit of the experimental spectra (Fig. 4a and 4c). Note that the reference spectrum of centre X in Figure 4d was measured on sample #3 on the same setup. Regardless of the nature of this extra spectral feature, an analysis of the $a$ and $b$ fitting parameters will be presented below. Note also that, even if nitrogen is not implanted, the vacancies produced by the oxygen implantation can form NV centres with the native nitrogen atoms during the annealing step and therefore be visible in the spectra at lower oxygen fluence, however in a smaller amount than for sample #1. Besides, the peak centred at 750 nm is attributed to neutral vacancies which are not completely annealed out despite the treatment at 1100 °C for 4 hours.

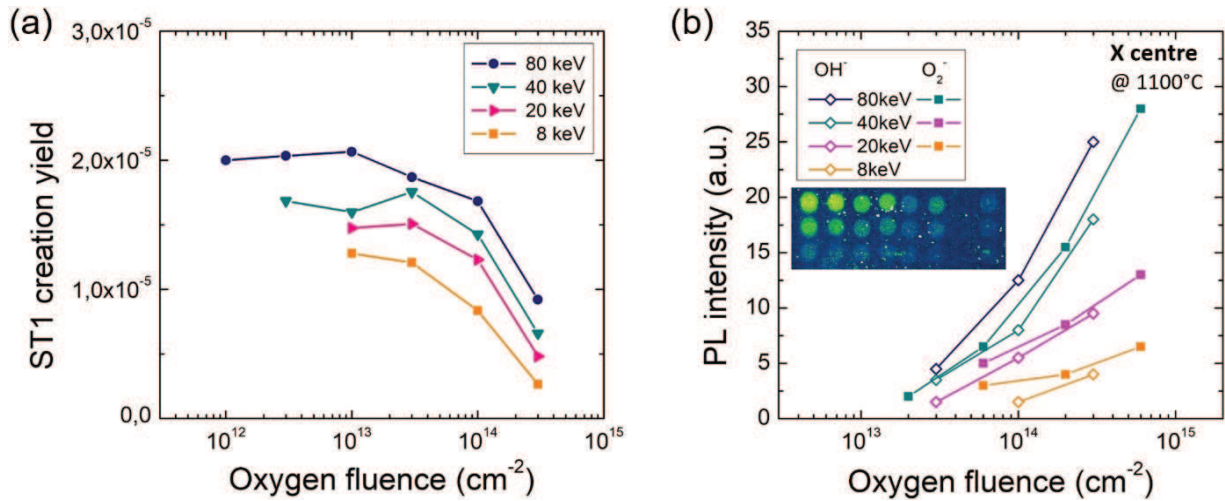

**Figure 5. ST1 and unknown X centres.** (a) Creation yield of the ST1 centre as a function of the implanted oxygen fluence, for the OH$^-$ ions of different energies after 1100°C annealing for 4 hours. (b) PL intensity of the unknown X centres as a function of the oxygen fluence (taking into account the factor of 2 between O$_2$ and O ions), for different ion energies and species. The inset shows the PL scan of the centres taken with 594 nm excitation and three filters (two 594 nm Notch filters and one 582 – 636 nm bandpass).

From the PL scans of Figure 3b and closer scans (not shown), the ST1 density can be extracted and compared to the implanted oxygen fluence to obtain the ST1 creation yield. This is plotted in Figure 5a as a function of the oxygen fluence, for the OH$^-$ species at different energies. Firstly, we observe that the creation yield is low compared to NV [25] or SiV [26] centres, i.e. between 1×10$^{-5}$ and 2×10$^{-5}$. This is in excellent agreement with previous works [15,16] also using 1100 °C annealing. It is particularly remarkable that such a low creation efficiency can be so reproducible upon several samples. Secondly, the ST1 yield starts to decrease when the oxygen fluence exceeds a few 10$^{13}$ cm$^{-2}$. This is also observed with implantation of NV centres [25]: their density saturates and subsequently decreases owing to a too large number of irradiation defects or to PL quenching when the ion fluence is further increased and approaches the amorphisation threshold of diamond. The formation of NVN (H3) centres also strengthens the decrease of NV PL by "consuming" the nitrogen to more preferentially form a defect centre different from the NV. Here, the oxygen is possibly "consumed" in a same way to produce other kinds of oxygen-related defects involving two atoms, as suggested in the first part, even if the temperature is here lower.

To illustrate the evolution of the PL spectra and the unknown spectral feature X, a PL map was recorded in the conditions where only the X centre is excited and emitting. To this scope, a 594 nm laser excitation was used, which does not (or extremely weakly) excite the ST1 centres. The detection band was set to 582 – 636 nm to collect a large part of this PL signal while avoiding collecting the emission from neutral vacancies (GR1) or NV$^-$ centres. Two Notch filters at 594 nm were also mounted to remove the laser signal. The PL map is shown in the inset of Figure 5b. The PL intensity was then extracted and plotted in Figure 5b as a function of the oxygen fluence. This clearly shows that the density of the X centre increases with the oxygen fluence, whereas the ST1 density drops.

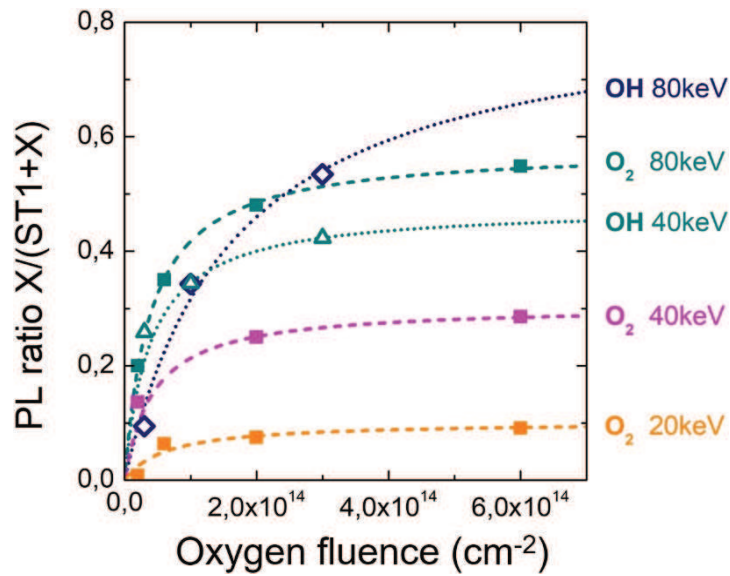

**Figure 6. Relative formation of ST1 and X defect centres.** The graph plots the relative presence of the X and ST1 centres (the ratio r = b/(a+b)) as a function of oxygen fluence, for different ion energies and ion species. The dots are experimental points obtained from the fit of PL spectra in Figure 4 with 532 nm excitation. The dotted lines are fits using equation (2). The fit parameters can be found in Table 1.

In order to quantify the relative occurrence of the ST1 and the unknown X centre in the sample as a function of the implanted oxygen fluence and of the implantation energy (and thus, depth), the $r = b/(a+b)$ ratio is plotted in Figure 6 as a function of the implanted oxygen fluence, using the fit parameters $a$ and $b$ from Equation (1). This ratio accounts for the relative weight of the unknown PL feature within the spectra. A threshold fluence for the efficient formation of the defect clearly emerges from this analysis. Interestingly, for each implantation energy, the part of the unknown centre X increases with the ion fluence, in good agreement with the data of Figure 5. This behaviour is still unclear, however the ratio $r$ can be suitably fitted using a "saturation" function $r_E(Fluence)$ in the form of:

$$r_E(Fluence) = r_0 \frac{1}{1 + F_{sat}/Fluence}$$

where $r_0$ represents the ratio at large fluence and $F_{sat}$ is the equivalent of a saturation fluence. The fit parameters $r_0$ and $F_{sat}$ used in Figure 4 can be found in Table 1. The strong dependence of the $r_0$ parameter on the implantation depth (i.e. ion energy) shows that there is a strong band bending effect for the X centre, either on its formation rate or on its charge state, limiting its formation close to the surface. There may therefore be room for improving the ST1 creation yield with respect to other species, especially using charge-state engineering.

| Implanted ion | Energy (keV) | $r_0$ | $F_{sat}$ (cm$^{-2}$) |
|---|---|---|---|
| OH$^-$ | 80 | 0.84 | $1.65 \times 10^{14}$ |
| OH$^-$ | 40 | 0.48 | $3.88 \times 10^{13}$ |
| O$_2^-$ | 80 | 0.58 | $3.90 \times 10^{13}$ |
| O$_2^-$ | 40 | 0.31 | $4.31 \times 10^{13}$ |
| O$_2^-$ | 20 | 0.10 | $6.34 \times 10^{13}$ |

**Table 1.** The parameters $r_0$ and $F_{sat}$ from equation (2) to fit the data in Figure 6.

**Cathodoluminescence spectroscopy (sample #3)**
Finally, cathodoluminescence measurements were conducted on sample #3, annealed at 1100 °C for 4 hours for different purposes. On the one hand, the excitation process involved in CL is very different from the PL one, and gives access to above-bandgap excitation and a wider detection range. On the other hand, different charge states may be emitting in CL in a significantly different way than in PL, as typically observed for NV centres [11]. Furthermore, CL spectroscopy may help to clarify the attribution of the X peaks observed in the previous spectra. To assess whether these peaks are oxygen-related or intrinsic, other ion species (silicon, carbon and nitrogen) were also implanted. The results are shown in Figure 7. The electron energy was set to 1 keV and a 100 pA electron current was adopted. All the spectra are background-corrected, spectrally corrected (from the detector sensitivity) and plotted in logarithmic scale and separated for clarity. At first, it is possible to identify the NV$^0$ ZPL (575 nm) as expected from CL measurements after nitrogen implantation [11] and the absence of NV$^-$ (638 nm) observed in PL, well illustrating the change of charge state that can occur under laser or electron beam excitation. The 389 nm emission from nitrogen-related defects and 503 nm emission from H3 centres (NVN) are also detected, with a lower intensity. For silicon implantation, the SiV$^-$ ZPL emission (738 nm) is present, even if rather weak if compared to PL (we do not have access to the 946 nm emission wavelength of the SiV$^0$).
Regarding oxygen implantation, the 535 nm ZPL of ST1 centres (in the only - yet unknown - charge state observed in PL up to now) is absent from the spectra, even for the highest oxygen fluence of $5 \times 10^{13}$ cm$^{-2}$. This shows that the ST1 charge state observed in PL is not active in CL excitation. The ST1 charge states remain to be elucidated but these results will likely support theoretical works in progressing on the structural attribution of the ST1 centre. It can however not be fully excluded that an efficient non-radiative recombination of charge carriers takes place on the majority oxygen-related centres (O$_S$, OV, OVH), whereas ST1 centres represent typically only a small fraction (i.e. 10$^{-5}$) of the implanted oxygen. We also remark a featureless broad band centred at 460 nm which is visible in the case of O$_2$ implantation (even for the lowest oxygen fluence of $5 \times 10^{11}$ cm$^{-2}$) but not for atomic O implantation or other species. The reason for this is not clear because the O$_2$ molecule with a kinetic energy of 80 keV is expected to break up while colliding the diamond surface, separating in two oxygen atoms implanted several nm apart. This should preclude the formation of any O$_2$V$_x$-like defect in sample #3 annealed at 1100 °C, whereas the CL results tend to suggest the formation of an O-related defect emitting at 460 nm.
In Figure 7a, it is striking that, regardless the implanted ion species, prominent sharp peaks at 550 nm, 562 nm, 580 nm and 594 nm are clearly visible, both in PL and CL spectroscopy. The X centre is therefore likely due to an intrinsic defect complex produced by the implantation and annealing processes. The spectral feature involves the two emission wavelengths noted as X centre in the beginning. We believe that they are belonging to the same defect. We add as a final remark that the X centre is not photostable. In PL, the emission intensity of this centre rapidly decreases, within a timescale of few seconds, under 600 µW excitation at 488 nm and seems to be weaker than in CL. In Figure 7b, two PL spectra of the high fluence carbon implantation ($3\times10^{14}$ cm$^{-2}$) annealed at 1100 °C are shown. On the top of the figure, the spectrum was acquired during 60 seconds while scanning over the implanted area and the peaks are very pronounced. At the bottom of the figure, the spectrum was instead acquired without moving and the fast photo-bleaching leads to a very different result where the X peaks are barely visible. It is worth noting that the PL spectra of sample #1 and #2 presented before in Figure 4 were taken by scanning the sample, therefore recording the maximum intensity of X centres before they photo-bleach. Finally, the relative CL intensity of the 460 nm broad peak was compared to the one of the X centre for the different implantation fluences (not shown) and the results are plotted in Figure 7c. Despite the bleaching behaviour of the X centre, the intensities of the 460 nm band and of the X centre are about the same order within the fluence range of two orders of magnitude. This indicates that the 460 nm band is indeed related to the $O_2$ implantation, as also suggested by its absence in the spectra of N and Si implanted areas.

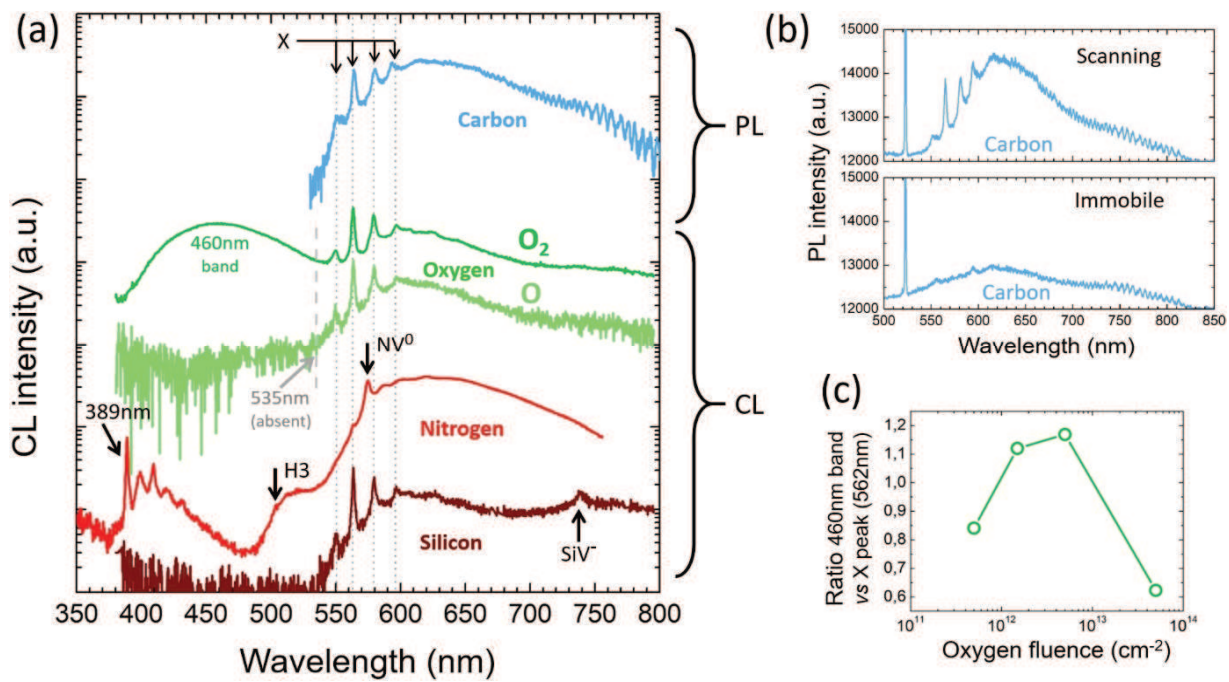

**Figure 7. Cathodoluminescence at 300K.** (a) CL and PL spectra of areas implanted with different ions species: Si$^-$ (60 keV, $1\times10^{12}$ cm$^{-2}$), CN$^-$ (75 keV, $1\times10^{12}$ cm$^{-2}$), O$^-$ (40 keV, $3\times10^{12}$ cm$^{-2}$), $O_2^-$ (80 keV, $5\times10^{12}$ cm$^{-2}$) and C$^-$ (40 keV, $3\times10^{14}$ cm$^{-2}$). (b) PL spectra of the carbon-implanted area, measured in different conditions. Top: the spectra is acquired for 60 s while the 488nm laser (0.6 mW) is scanned across the area. Bottom: the spectra is acquired while the laser is kept at the same place. (c) Comparison of the CL peak intensity of the 460 nm band and of the X centre in the case of $O_2$ implantation at 80 keV for different fluences.

**Conclusions**

In this work, we investigate using photo- and cathodoluminescence spectroscopy the formation and stability of oxygen-related centres in diamond created by oxygen ion implantation and annealing at different temperatures. We accurately identify the temperature stability range of the ST1 centres, which have a maximum creation efficiency at with an annealing process at 1100 °C. Contrarily to NV centres or other well-known impurity-vacancy colour centres in diamond, they are not yet formed at 900 °C. This result provides insight into the ST1 formation energy and competing processes during thermal annealing which will likely help identify the still unknown structure of the ST1 centre. For a 1100°C annealing, the ST1 ensembles show a ZPL at 535 nm and well-contrasted spin resonance spectra for the $|0\rangle \leftrightarrow |-\rangle$ and $|0\rangle \leftrightarrow |+\rangle$ transitions within the metastable triplet state, well suitable for magnetic field sensing by ODMR. The zero-field splitting $D$ is of 1133.5 MHz and the crystal field $E$ of 143.5 MHz, at room temperature. The ST1 creation yield shows a slight variation with the implantation energy and is in agreement with previous works. Despite its low value of about $10^{-5}$ per oxygen atom, ST1 centres can be created with high reproducibility. Our PL results and the low value of ST1 creation yield indirectly suggest that the more likely O-species formed in diamond such as $O_S$, OV and OVH are not optically active. Interestingly, the ST1 centre (in its charge state emitting at 535 nm) could not be excited by CL. Besides, CL measurements show a broad band at 460 nm created in the case of $O_2$ implantation but not atomic oxygen. It is however still unclear whether the band is related to an oxygen defect and further investigations are necessary to clarify this aspect. For a higher annealing temperature of 1500 °C, another defect centre is observed, with characteristic lines at 584.5 nm and 598.5 nm, labelled Y and tentatively attributed an $O_2V_x$ configuration. It is only formed for ion fluences larger than $1\times10^{13}$ cm$^{-2}$ and its emission intensity does not increase linearly with the ion fluence but rather following a power law characterised by an exponent comprised between 1.5 and 1.9. This suggests a defect containing two oxygen atoms, formed by oxygen diffusion, similarly to the NVN (H3) defect formation at high nitrogen density and high temperature treatment. It was however not possible to isolate individual centres. This indicates that their quantum efficiency is rather poor and/or that they preferentially decay using non-radiative channels. Finally, our results also reveal the creation of another defect with a broad PL emission and strong narrow lines at 550 nm, 562 nm, 580 nm and 594 nm, still unknown, but of intrinsic nature because it is formed whatever ion species is used. This centre, named here X, efficiently forms for high ion fluences typically larger than $3\times10^{13}$ cm$^{-2}$ and with an intensity increasing with the implantation depth. It is therefore likely sensitive to surface band bending and is not photostable. To conclude, the atomic structures of the ST1 and Y oxygen-related centres still need to be correctly attributed, but their thermal formation and stability estimated in this work represent valuable data for their reproducible creation and their identification by means of *ab initio* methods. Indeed, the calculation of the formation energy of a given defect can be obtained much faster than its electronic and spin properties.

**Methods**
**Ion implantation and sample preparation**
Three diamond samples of type IIa and (001) surface were used for this study. Sample #1 was grown at LSPM by CVD on a HPHT substrate, following a standard growth procedure for non-intentionally doped diamond. The CVD layer has a thickness of xxx μm and was neither polished nor treated after the growth and before the ion implantation. Samples #2 and #3 are commercially available diamonds from Element6 grown by CVD and polished. The specifications certify a nitrogen concentration of less than 5 ppb.
Two different ion implantation setups were used. Sample #1 was implanted using an Orsay Physics focused ion beam (FIB) column at IRCP. The FIB is a gas source column with the ability to implant ions over an energy range of 3 – 30 keV. The column contains a Wien filter allowing the selection of specific ions species. In the case of this work, $O^+$ and $O_2^+$ ions with energies of 30 and 10 keV were implanted.
Samples #2 and #3 were implanted using the 100 kV accelerator at the University Leipzig. This accelerator produces negative ion species such as $O^-$, $OH^-$ or $O_2^-$. The home-made source cathode was pressed from a powder of iron oxide. After extraction, the mass separation of the ion beam is then ensured by a 90° bending electromagnet. Both molecular and atomic oxygen were implanted. Note that the hydrogen contained in the OH penetrates much deeper than the oxygen atom and its effect is neglected. Hydrogen is anyway already present as the main impurity in CVD ultrapure diamond. The ion beam was collimated by a 50μm aperture and several circular areas were implanted in the energy range 8 – 80 keV, with fluences ranging from $1\times10^{12}$ to $3\times10^{14}$ cm$^{-2}$.

**PL and CL imaging and spectroscopy and spin measurements**
The optical measurements were conducted at room-temperature on a home-made confocal fluorescence microscopy setup equipped with an air objective with NA = 0.95 and with single-photon counting modules. Three different lasers were used (488, 532 and 594 nm) to preferentially excite either one or the other defect centres. Different filters were employed to select the spectral range in detection.
For the spin measurements (ODMR), a copper wire was soldered on the sample and placed in the vicinity of the implanted area.

Cathodoluminescence spectra were measured at room-temperature using a IT800 JEOL scanning electron microscope coupled with a MONARC PRO GATAN cathodoluminescence system. A parabolic mirror inserted into the SEM chamber efficiently collects the light from the samples located 1 mm below. Spectra are then acquired in a single pass by the integrated spectrometer using a 300 groove per mm grating and a PIXIS 400B cooled Charge Coupled Device (CCD).

**CRediT authorship contribution statement**
**Paul Neugebauer:** Data curation, Investigation, Resources. **Xinxi Huang:** Data curation, Investigation, Writing – review & editing. **Chloe Newsom:** Resources, Writing – review & editing. **Christophe Arnold:** Investigation, Writing – review & editing. **Hjørdis Martelock:** Investigation, Writing – review & editing. **Séverine Diziain:** Investigation, Writing – review & editing. **Edoardo Monnetti:** Resources. **Jocelyn Achard:** Resources, Funding acquisition, Writing – review & editing. **Tobias Lühmann:** Resources, Writing – review & editing. **Paolo Olivero:** Project administration, Writing – review & editing. **Jan Meijer:** Project administration, Resources, Funding acquisition. **Julien Barjon:** Investigation, Funding acquisition, Writing – review & editing. **Alexandre Tallaire:** Resources, Funding acquisition, Writing – review & editing. **Sébastien Pezzagna:** Conceptualisation, Data curation, Investigation, Supervision, Writing – original draft, Writing – review & editing.

**Declaration of competing interest**
The authors have no financial or non-financial conflicts of interest to disclose. The authors confirm that this work is original and has not been published elsewhere, nor is it currently under consideration for publication elsewhere. No artificial intelligence tools or services were used to write, construct or organise any part of this manuscript.

**Acknowledgements**
This work has received funding from the German Federal Ministery of Research, Technology and Space (project CoGeQ, Grant No. 13N16097), from ESR/EquipEx + program (Grant No. ANR-21-ESRE-0031) and has been supported by Region Ile-de-France in the framework of DIM QuanTiP.